\documentclass[superscriptaddress,groupedaddress,nofootnoteinbib,11pt]{article}
\pdfoutput=1
\usepackage{color}
\usepackage{graphicx}
\usepackage{dcolumn}
\usepackage{bm}
\usepackage{amssymb}
\usepackage{amsmath}
\usepackage{sectsty}
\usepackage{colortbl}

\usepackage{latexsym}
\usepackage{float}
\usepackage{ifthen}
\usepackage{enumerate}
\usepackage{url}
\usepackage[force]{feynmp-auto}
\usepackage{jcappub}
    \usepackage{picinpar}
    \usepackage{colortbl}
\usepackage{multirow}
	\usepackage{float}
	      \usepackage{setspace}
\usepackage{array}
\usepackage{bm}
\usepackage{amsopn}
\usepackage{tablefootnote}
\renewcommand{\vec}[1]{\bm{\mathrm{{#1}}}}
\usepackage{booktabs}
\usepackage[table]{xcolor}
\definecolor{lightgray}{gray}{0.9}

\newcommand{\p}{\partial}

\def\ba{\begin{eqnarray}}
\def\ea{\end{eqnarray}}
\def\beq{\begin{eqnarray}}
\def\eeq{\end{eqnarray}}
\def\noi{\noindent}

\def\({\left(}
\def\){\right)}

\def\p{\partial}

\def\<{\langle}
\def\>{\rangle}

\definecolor{verde}{rgb}{0,0.5,0}

\usepackage{titlesec}

 \def\bea  {\begin{eqnarray}}   \def\eea  {\end{eqnarray}}
  
\def\P{{\cal P}}

\def\S{{\cal S}}
\def\de{{\delta}}

\def\H{{\cal H}}
\def\la{{\langle}}
\def\ra{{\rangle}}

\definecolor{verde}{rgb}{0,0.5,0}

\def\be{\begin{equation}}
\def\ee{\end{equation}}
\def\bea{\begin{eqnarray}}
\def\eea{\end{eqnarray}}
\def\be{\begin{equation}}
\def\ee{\end{equation}}
\def\ba{\begin{align}}
\def\ea{\end{align}}
\def\p{\partial}
\def\noi{\noindent}

\renewcommand\({\left(}
\renewcommand\){\right)}

\newcommand\lsim{\mathrel{\rlap{\lower4pt\hbox{\hskip0.5pt$\sim$}}
    \raise1pt\hbox{$<$}}}
\newcommand\gsim{\mathrel{\rlap{\lower4pt\hbox{\hskip0.5pt$\sim$}}
    \raise1pt\hbox{$>$}}}

\newboolean{editorial}
\setboolean{editorial}{true}
\newcommand{\editorial}[2]{\ifthenelse{\boolean{editorial}}

{\textcolor{red}{ [\textsf{\textbf{{#1}}}:} \textcolor{blue}{\textsf{{#2}}}\textcolor{blue}{]}}{}}

\definecolor{dullpurple}{rgb}{0.431,0.188,0.534}
\definecolor{darkgreen}{rgb}{0.133,0.545,0.133}
\definecolor{verde}{rgb}{0,0.5,0}

\newboolean{comment}
\setboolean{comment}{true}
\newcommand{\comment}[2]{\ifthenelse{\boolean{comment}}{\textcolor{blue}{{{{#1}}}: }\textcolor{verde}{{#2}}}{}\\}

\newboolean{red2}
\setboolean{red2}{true}
\newcommand{\redtwo}[2]{\ifthenelse{\boolean{red2}}{\textcolor{red}{#1}}\\}

\begin{document}

\title{On Observables in a Dark Matter-Clustering Quintessence System }

\author{Matteo Fasiello$^{a,c}$ and Zvonimir Vlah$^{b,c,d}$}
\affiliation{$^a$Institute of Cosmology and Gravitation, University of Portsmouth, PO1 3FX, UK.}
\affiliation{$^b$Theoretical Physics Department, CERN, Geneva, Switzerland.}
\affiliation{$^c$Stanford Institute for Theoretical Physics  Department of Physics, Stanford University, Stanford, CA 94306}
\affiliation{$^d$Kavli Institute for Particle Astrophysics  Cosmology, Stanford University and SLAC, Menlo Park, CA 94025.}

\abstract{
We consider a system where dark matter dynamics is enriched by the presence of clustering quintessence in the approximation where the system is effectively reduced to one degree of freedom. We study the corresponding observables up to one-loop order and then point out similarities between the power spectrum of the reduced system and the behaviour of non-equal time pure dark matter correlators.
We then focus on the one-loop total density power spectrum in the IR limit as a diagnostic tool for consistency relations breaking. Unlike the non-equal time case, the reduced system does still obey consistency relations;  we illustrate this by explicitly verifying the 1-loop IR cancellation. A more general setup, obtained by relaxing the assumption of a vanishing sound speed, is also analyzed. In this and similar scenarios the presence of additional dynamics, typical of dark energy and modified gravity models, implies that one may no longer gauge away the  squeezed contribution of observables such as the dark matter bispectrum. We show how these effects propagate all the way to biased tracers.
}

\maketitle

\section {Introduction}
\label{section0}

\noi A detailed understanding of the mildly-non-linear formation of structure in the universe is of  paramount importance for cosmology. These scales carry crucial information both on early (e.g. non-Gaussianities \cite{Dalal:2007cu}) and late-time (e.g. current cosmic acceleration \cite{Huterer:2013xky}) physics. Several approaches have been developed to tackle the dynamics in these regimes, with N-body simulations as the one able to probe deeper into the highly non-linear regime.\\
Remarkably, it is also possible to construct consistent perturbative formalisms \cite{Bouchet:1995ez}-\cite{Carrasco:2013mua} that deliver observables up to scales of about $\sim 0.5\,[{\rm h/Mpc]}$. The perturbative framework, although severely limited in k-reach, enables one to account for the various layers of physics that make up actual large scale structure observables. From baryonic effects \cite{Lewandowski:2014rca} to biased tracers\cite{Senatore:2014eva} to redshift space \cite{Lewandowski:2015ziq,delaBella:2018fdb}; from primordial non-Gaussianities \cite{Angulo:2015eqa} to the extra degrees of freedom of dark energy and modified gravity models. In particular, the effective field theory treatment \cite{Baumann:2010tm} of large scale structure encapsulates the action of unaccessible smaller scales on (at reach) long-wavelength modes in a number of ``UV" coefficients, to be determined by comparison with simulations and, in the near future, observations.\\ 
In what follows we adopt the perturbative treatment to investigate the effects of adding a clustering quintessence component to cold dark matter in the fluid description. Such set-up enjoys drastic simplifications in the limit of vanishing quintessence sound speed. The resulting dynamics serves as an illustrative proxy for systems that go beyond the $\Lambda$CDM paradigm. \\
We find that the reduced ($c_s\rightarrow 0$) system still enjoys properties typical of the pure dark matter case, such as so-called consistency relations between the equal-time squeezed bispectrum and the power spectrum. The same properties are behind the well-known 1-loop IR cancellation, which we verify explicitly for the simplified system. A non-zero sound speed, on the other hand, signals the presence of a non-adiabatic mode which may break consistency relations.\\   

\noi This paper is organized as follows: in \textit{Section} \ref{section1} we first illustrate the dynamics of the reduced system and then provide a parallel with non-equal-time correlators; we show why the observables are similar and elucidate where the parallel breaks down; in \textit{Section} \ref{consistency} we provide a brief general treatment of consistency relations in large scale structure set-ups, with special focus on the case when a dark energy or a modified gravity component is present; in \textit{Section} \ref{cbt} we show how these effects propagate to biased tracers.

\section {Total power spectrum}
\label{section1}
\begin{figure}
\qquad \qquad\qquad\qquad
\includegraphics[scale=0.49]{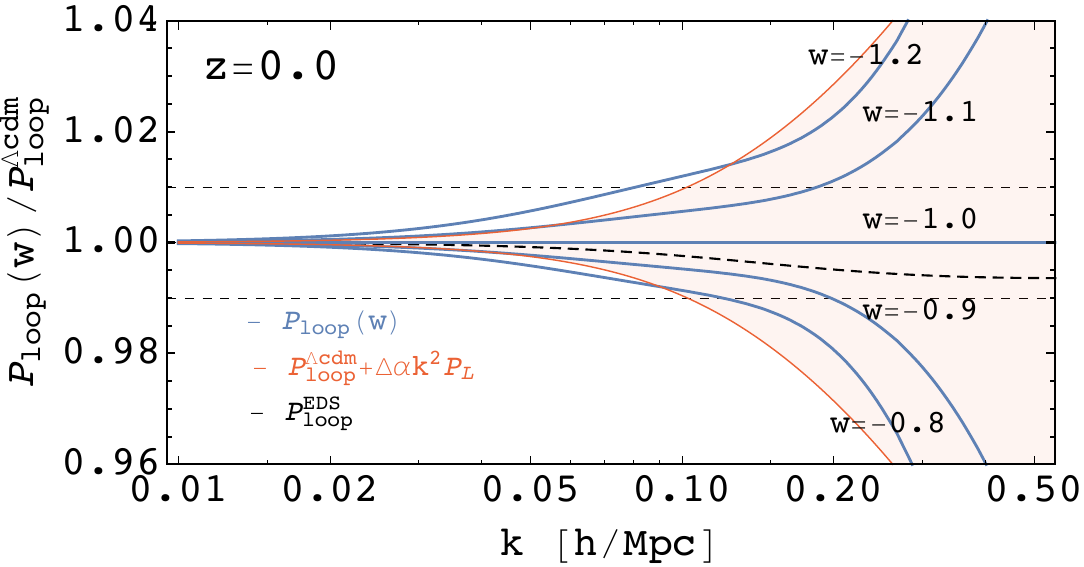}
\caption{The total power spectrum as a function of the wavenumber normalized w.r.t. the $\Lambda$CDM result. It is clear how the clustering effect emerges only at midly-non-linear scales. In orange the correction due to an EFT counter-term with the typical $(k/k_{\rm NL})^2$ scaling. The dashed black line is obtained using the EdS approximation.} 
\label{fig0}
\end{figure}

\noi Additional degrees of freedom (e.g. those of dynamical dark energy) can be added to the fluid description of dark matter, thereby generating a system of {gravitationally} coupled equations. In the next section, we will describe some of the properties of such a system. In this one we study the simplified dynamics of a clustering quintessence model in the vanishing sound speed approximation first analyzed in \cite{Sefusatti:2011cm}. The continuity and Euler equation read 
\begin{align}
 \quad \p_\tau \delta_T + \partial_i [(C+ \delta_T)v^i ] =0\; , \qquad 
\p_\tau v^i+ \mathcal{H}v^i+v^j \p_j v^i  =-\nabla^i  \Phi \;;
\label{packaged}
\end{align}
where the total density contrast $\delta_T$ has been defined as the combination of dark matter and quintessence densities weighted by density parameters as in $\delta_T\equiv \delta_m +\frac{\Omega_Q}{\Omega_m}\delta_Q $. The variables $\Phi$ and $v^i$ stand respectively for the gravitational potential and the velocity field. We stress in particular the time-dependent quantity defined as
\bea C(\tau)=1+(1+w)\frac{\Omega_Q}{\Omega_m}(\tau)\; ,
\eea
where the latter is equal to unity in $\Lambda$CDM.  
The system is closed by Poisson's equation $\nabla^2 \Phi = \frac{3}{2}\mathcal{H}^2\Omega_m \delta_T$. The dynamics can be solved perturbatively. The results for the tree-level bispectrum are given in \cite{Sefusatti:2011cm} whilst the one-loop power spectrum, as well as all-order integral solutions for the fields, were found in \cite{Fasiello:2016qpn}.
The kernels for the total density fluctuations $\delta_T$ up to the third order are\begin{align}
F_{2}&= -\tfrac{1}{2} \left ( 1 - \epsilon^{(1)}- \tfrac{3}{2}\nu_2 \right )\alpha_s + \tfrac{3}{2} \left ( 1 - \epsilon^{(1)} - \tfrac{1}{2}\nu_2 \right) \beta, \nonumber \\
F_{3}&= (1-\epsilon^{(2)}) \mathcal{F}_3^{\epsilon}+\nu_3 \mathcal{F}_3^{\nu_3}+  (1-\epsilon^{(1)}) \nu_2 \mathcal{F}_3^{\nu_2} + \lambda_1 \mathcal{F}_3^{\lambda_1} +\lambda_2 \mathcal{F}_3^{\lambda_2}\,, 
\label{eq:kernels}
\end{align}
where for simplicity we have suppressed momentum-vector dependence as well as time dependence in $\epsilon, \nu_3, \nu_2,  \lambda_2$ and $\lambda_1$ (for explicit definitions we refer the reader to the Appendix in \ref{appe}; for a derivation see instead \cite{Sefusatti:2011cm,Fasiello:2016qpn}). The reduced kernels such as $\alpha$, $\beta$, and $\mathcal{F}_3^{\epsilon}$ are time-independent and are only function of  momenta \cite{Fasiello:2016qpn}.
We reproduce here the explicit form of the quantities $\epsilon^{(n)}$ since they will be of particular importance in what follows. These read:
\bea 
\epsilon^{(1)}(\eta)&=&1-e^{-\eta} \int^{\eta}_{-\infty}d\tilde{\eta}[e^{\tilde{\eta}}/C(\tilde{\eta})]\nonumber \\ \epsilon^{(2)}(\eta)&=& 2\int_{-\infty}^{\eta}d\tilde{\eta} ~e^{2(\tilde{\eta}-\eta)} \left( 1-(1-\epsilon)/C(\tilde{\eta}) \right) \; .
\eea
\noi Note that, by construction, both vanish in the simplified case where $C=1$, as the system in Eq.~\eqref{packaged} reduces to the pure dark matter case. We consider the one-loop power spectrum for the total density:
\begin{equation}
P_{\rm 1-loop}(k,a) = P_{\rm L}(k,a) + P_{22}(k,a)+2P_{13}(k,a) + P_{\rm c.t.}(k,a),
\end{equation}
where each of the above contributions is defined as
\begin{align}
P_{{\rm L},k}(a) &= D^2_+(a) P^{\rm in}_{\bf k},\; \nonumber\\
P_{22,k}(a) &= 2D^4_+(a) \int_{\bf q} \big[ F_{2}({\bf k} -{\bf q}, {\bf q},a) \big]^2 P^{\rm in}_{|{\bf k} - {\bf q}|} P^{\rm in}_{\bf q} \, , \nonumber\\
P_{13,k}(a) &= 3D^4_+(a) P^{\rm in}_{\bf k} \int_{\bf q} F_{3}({\bf k}, -{\bf q}, {\bf q},a)  P^{\rm in}_{\bf q}\, .
\end{align}
Here $D$ is the linear growth function and $P_{\rm c.t.}(k,a)$ stands for the one-loop \textit{counterterm}, encoding short-scale dynamics. It is given in the \textit{EFT of LSS} 
\cite{Carrasco:2012cv} {simply as $\propto k^2/k^2_{\rm NL}\,P_{\rm L}$}, and multiplies a to-be-determined (by, for example,  comparison with N-body simulations) numerical coefficient.
$P^{\rm in}_{k}$ is the time-independent initial power spectrum obtained form Boltzmann algorithms such as \cite{Lewis:1999bs,Lesgourgues:2011re}.\\
\noi We present here for the first time several plots of the total power spectrum of the reduced clustering quintessence system for the $w \not=1$ case. The comparison in Fig.~(\ref{fig0}) with the $\Lambda$CDM result underscores how the clustering quintessence system deviates from standard behaviour in the high-$k$ regime. The reason becomes clear after repackaging Eq.~(\ref{packaged}) as 
\bea
&&\frac{\partial \delta_{\bf k}}{\partial\eta}-\Theta_{\bf k}=\frac{\alpha({\bf q}_1,{\bf q}_2)}{C(\eta)} \Theta_{\bf q_1} \delta{\bf q_2}\qquad\nonumber\\
&&\frac{\partial \Theta_{\bf k}}{\partial\eta}-\Theta_{\bf k}
-\frac{f_{-}}{f^2_{+}} (\Theta_{\bf k}-\delta_{\bf k}) =\frac{\beta({\bf q}_1,{\bf q}_2)}{C(\eta)} \Theta_{\bf q_1} \Theta{\bf q_2}\;, \qquad\;
\label{repack}
\eea
where the following definitions have been used
\bea
f^{+/-}\equiv \frac{d\,{\rm ln} D^{+/-}}{d\, {\rm ln} a}\,; \;\; \theta_{\bf k}\equiv D_{+} \delta^{\rm in}_{\bf k}\,;\;\; \Theta_{\bf k}\equiv -\frac{C}{\mathcal{H}f_{+}}\theta_{\bf k}\, ,\quad 
\eea
with $D^{+/-}$ as the solutions for the linear growth rate, and $\alpha({\bf q_1},{\bf q_2}),\beta({\bf q_1},{\bf q_2})$ the standard (see e.g. \cite{Bernardeau:2001qr}) kernels . The non-trivial time dependence of $C(\tau)$ appears in Eq.~(\ref{repack}) only on the righ hand side, that is only at non-linear order in perturbation theory. Observations at those scales dictate that these extra effects must remain very small (i.e. at most a few percent level). Notice that this is in direct contradistinction to the case of screened dark-energy and modified gravity theories. There, above a (model-dependent \cite{Fasiello:2017bot}) threshold value for the momentum, the effective coupling between dark matter and any additional degree of freedom is suppressed and the system flows back to $\Lambda$CDM. This screening behaviour is characteristic of models where a so-called fifth-force is present: (i) at linear scales an order one difference is allowed for observables with respect to their cold dark matter + cosmological constant counterpart; (ii) conversely, the dynamics at smaller scales ought to be unaffected by any fifth force in order to recover general relativity.

\begin{figure}[H]
\qquad \qquad\qquad \qquad \includegraphics[scale=0.51]{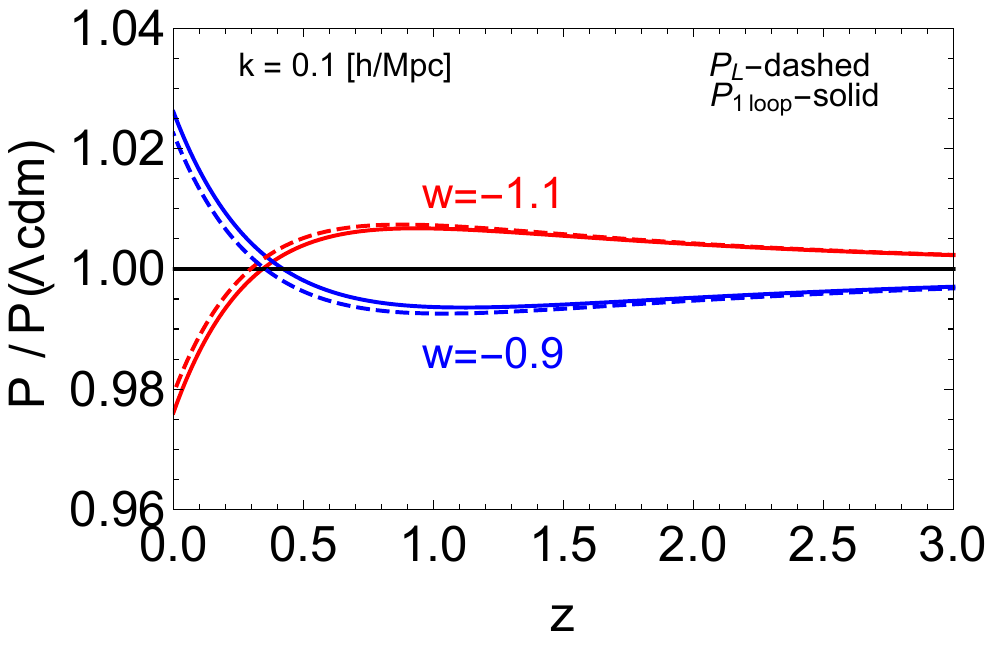}
\caption{The total power spectrum as a function of the redshift $z$ for different values of the equation of state parameter $w$. Dashed lines correspond to the linear result, solid curves to the observable including the one loop contribution.} 
\label{fig02}
\end{figure}

\noi It is also interesting to plot (see Fig.~\ref{fig02}) the power spectrum, up to one loop, as a function of redshift. We have normalized it so as to match the pure dark matter result at early times, where $D(\tau)\simeq a(\tau)$. 

\noi Of course, although upcoming surveys will soon allow us to probe increasingly large redshifts, the observables at our disposal are typically biased tracers whose description demands we include several layers of additional dynamics. We will return to this point in \textit{Section} \ref{cbt}.

\subsection*{Parallel with non-equal time correlators}
\begin{figure*}[t!]
\includegraphics[scale=0.58]{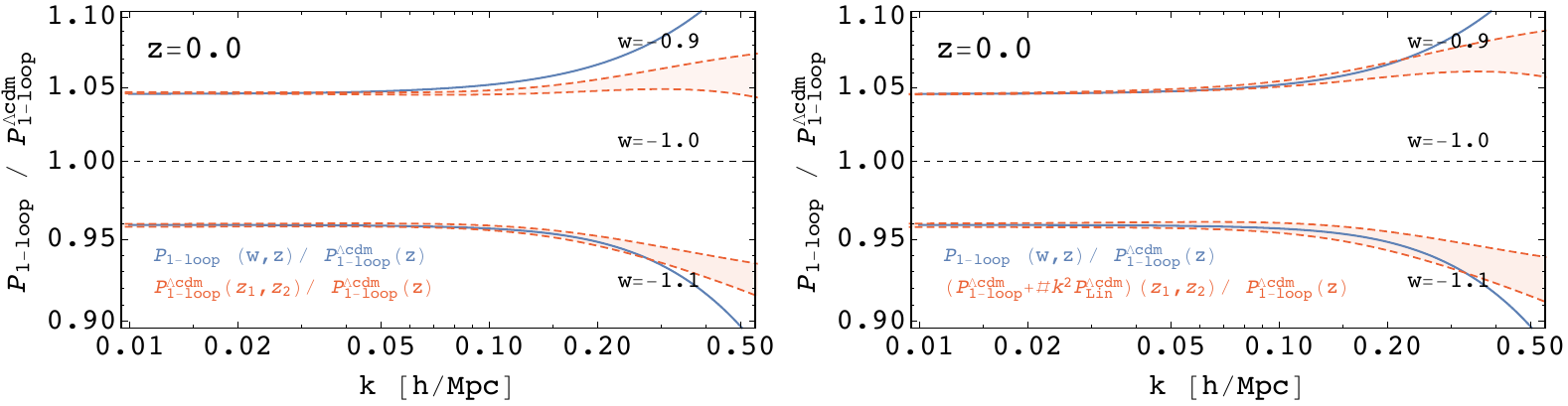}
\caption{The total DM+Q density power spectrum for $w=-1.1$ and $-0.9$, normalized by the $\Lambda$CDM ($w=-1.0$) power spectrum (for the latter, we use an Einstein-de Sitter-type approximation, i.e. we neglect the time dependence of the kernels, which is standard practice in the field), is shown up to one-loop order in solid blue lines. 
 \textit{Left:} the red dashed lines and the red band represent the approximation obtained from the non-equal time $\Lambda$CDM power spectrum choosing $z_1$ and $z_2$ different from the nominal $z$. This power spectrum too is normalized by the $\Lambda$CDM power spectrum at the nominal redshift $z$. 
 \textit{Right:} In addition to the non-equal time $\Lambda$CDM power spectrum, red dashed lines and the red band contain an additional small contributions $ \sim k^2P_{\rm lin}$. This corresponds to the small change of value of the EFT parameter (counterterm), which can 
 further extend the range of validity for this approximation.} 
\label{fig1}
\end{figure*}

\noi We elaborate in this section on an intriguing fact: by judiciously choosing two different redshifts for the ${\rm \Lambda CDM}$ case, the DM+quintessence and ${\rm \Lambda CDM}$ systems can provide consistently similar contributions for observables such as the power spectrum up to at least one-loop order. In Fig.~(\ref{fig1}) we plot the the total (DM+quintessence) power spectrum results up to one-loop for different values of $w$, at redshift $z=0$, as a function of momentum. In the same plots we show the non-equal time power spectrum for the $\Lambda$CDM case up to one-loop. Both power spectra are normalized by the same ${\rm \Lambda CDM}$ power spectrum quantity. 
From the plots in Fig.(\ref{fig1}), one can see that {the two power spectra} are very similar ($\ll1\%$ difference) 
on scales where the one-loop results are expected to be valid, 
$k \lesssim 0.15 \rm{Mpc}/h$  \cite{Carrasco:2012cv}. On the right panel of Fig.(\ref{fig1}) we add a typical (in the sense of the ``EFT of LSS" approach \cite{Baumann:2010tm}) counterterm contribution. {This shows that a small 
change in the counterterm parameter values assures that an unequal-time power spectrum can mimic our DM+quintessence {10\%} deviations from (equal time) $\Lambda$CDM} up to and beyond $k \lesssim 0.2 \rm{Mpc}/h$. \\
\noi {  
The similarities between the power spectrum of the DM-quintessence system and that of non-equal time DM fields stems from the fact that a non-trivial time-dependence in $C(\tau)$ acts as an additional ``clock" in the DM-only dynamics. As such, it mimics non-equal time-dependent observables. This correspondence is in place for the full power spectra starting from the linear contributions until up to, and possibly further than, one-loop. We find this rather intriguing and, as a consequence, it is worthwhile to explore how far we can take this parallel. In order to do so, we will employ a crucial probe of extra dynamics: the breaking of so-called consistency relations (CRs).\\
The latter are specific relations between observables that stem from (residual) symmetries in the description of the physical system at hand. We defer a discussion of consistency relations to the next section and focus here on the consequences of one of the symmetries behind CRs in large scale structure: \textit{Galilean invariance}.}\\ In $\Lambda$CDM, it is well-known that a cancellation between the leading infrared (IR) contributions occurs { for equal time correlators} {\cite{Jain:1993jh,Scoccimarro:1995if,Kehagias:2013yd,Peloso:2013zw,Blas:2013bpa}}. It was recognized already in \cite{Jain:1993jh} that Galilean invariance (invariance under time-dependent translations) is at the heart   of the cancellation. Later works have further investigated this result and placed it squarely in the context of consistency relations.\\ It is expected that extra dynamics, such at that due to the presence of extra non-adiabatic degrees of freedom, will break consistency relations already for equal-time correlators with Gaussian initial conditions. Things are a little more subtle for the system of Eq.~(\ref{repack}): although the equations appear to describe one degree of freedom, there is ``remnant" of its two-species origin in the time dependence of $C(\eta)$. On the other hand, the two species only interact gravitationally and the only density constrast we can probe is ultimately what appears on the right hand side of Poisson's equation, namely $\delta_T$. We ask then:
does the breaking occur in the case of our reduced system? Is this the reason behind the similarity  between the reduced system power spectrum and the non-equal time $\Lambda$CDM two-point correlator?\\
\noi Let us verify, by a direct 1-loop order calculation, 
whether the IR cancellation is still in place. For the leading infrared contributions we find
\begin{align}
\frac{p^{{\rm 1-loop}}_{k, q}(a)}{P_{{\rm L},k}(a) P_{{\rm L},q}(a)}\Big|_{{k \gg q}} \sim \left( \left(1-\epsilon^{(1)} \right)^2 - 1 + \epsilon^{(2)} \right)  \frac{2}{3} \frac{k^2}{q^2} ,
\label{55}
\end{align}
where $p^{{\rm 1-loop}}_{k, q}$ is the integrand of the one-loop power spectrum: $P^{{\rm 1-loop}}_{k} = \int_{\bf q} p^{{\rm 1-loop}}_{k, q}$.
For $C(\tau)=1$ (i.e. in the $\Lambda$CDM limit) 
as well as for a generic constant $C$ we naturally recover the expected cancellation. The non-trivial information comes from the fact that, even for a time-dependent $C(\tau)$, the system still enjoys the IR cancellation (up to higher order corrections). This can be readily verified after some algebra upon direct substitution of the $\epsilon^{(n)}$ functions in Eq.(\ref{55}). Crucially, this result is in contradistinction to the \textit{non}-equal time $\Lambda$CDM dark matter power spectrum. Let us show this at the level of the the same IR limit as in Eq.(\ref{55}):
 \begin{align}
\frac{p_{{\rm 1-loop},k, q}(a_1,a_2)}{P_{{\rm L},k}(a_1) P_{{\rm L},q}(a_2)}\Big|_{{k \gg q}} \sim -\frac{\left( D(a_1) - D(a_2) \right)^2}{D(a_1) D(a_2)}  \frac{1}{3} \frac{k^2}{q^2} ~,
\label{6}
\end{align}
where $D$ is the linear growth function.\\
\noi We must then conclude that the one between these two systems, even when limited up to one loop, is an intriguing parallel but is far from being an exact equivalence. We have been focussing on the 1-loop IR limit because of its importance for the squeezed configuration of the corresponding observables: this is the limit in which so-called consistency relations are relevant.
In turn, consistency relations are relevant as a natural probe of extra dynamics in a physical system.\\
Our findings underscore that the main contributions to the observables plotted in Fig.(\ref{fig1}) do not originate from the IR limit in, respectively, Eq.(\ref{55}) and (\ref{6}). In addition to the IR contributions, one must indeed account for the linear piece as well as the contributions from other 1-loop configurations. These, depending on $k$, will be the leading ones. Having verified, by means if the IR cancellation, that consistency relations are still active for the reduced system, we now expand the analysis to the full system, allowing a non-zero sound speed for the quintessence component.

\section {Consistency relations for clustering quintessence}
\label{consistency}
\noi In this section we shall adopt the notation of \cite{Assassi:2012zq},\cite{Horn:2014rta}. Consistency relations (hereafter CRs) stem from a residual gauge symmetry of the action or the equations of motion (eom) of a physical system. Although certain gauges, such as unitary gauge, are known to completely fix  diffeomorphism (diff) invariance, the fact that such fixing is complete is strictly true only for diffs that vanish at spatial infinity. Indeed, the residual gauge symmetry CRs rely upon is that of diffs that do not vanish at infinity \cite{Weinberg:2003sw}.\\
One may derive non-trivial CRs when the soft mode characterizing  any squeezed limit transforms non-linearly under the residual diff. In the context of large scale structure, using the fluid treatment for pure dark matter dynamics, one may show that the system equations possess a time dependent symmetry under which the velocity potential $\pi$ and the Newtonian potential $\Phi$ transform non-linearly \cite{Kehagias:2013yd,Peloso:2013zw},\cite{Horn:2014rta}. It follows that the effect of a long mode $\pi_L$ on $n$ short modes corresponds to the action of a residual gauge-symmetry on the observable made up by the corresponding n-point function and can, as such, be gauged away. This often translates into a suppressed signal for the squeezed n+1-correlator. \\
The mere requirement that the eom are invariant under a residual diff is not enough to guarantee that the CRs (at least in their standard formulation) are in place. Following e.g. \cite{Horn:2014rta} one may list three main necessary conditions:\\
- Symmetry of the action (eom) under the residual diff;\\
- Single-clockness: the transformation of an array of $n$ hard modes is mapped to the presence of \textit{one} soft mode;\\
- Adiabaticity: the eom of the gauge parameter describing the residual diff ought to mimic that of a long physical mode.\\ 
Our reduced system satisfies these requirements.
To clarify the picture and set the stage for generalizations, we report the full system in Eq.(\ref{fullm}-\ref{fullP}) below, in the approximation already in use in \cite{Sefusatti:2011cm}, \cite{Anselmi:2011ef} and in the Newtonian limit. For dark matter, we have
\begin{align}
\label{fullm}
\frac{\p \delta_m}{\p \tau}+\p_i [(1+\delta_m) v_m^i]=0\; ,\quad 
\frac{\p v_m^i}{\p \tau}+ \mathcal{H}v_m^i+v_m^j \p_j v_m^i  =-\nabla^i  \Phi \; ,  \nonumber\\ 
\end{align}
that is, the standard continuity and Euler equations. Lifting the $c_s\rightarrow 0$ limit, the (linear) quintessence equations are instead modified, to give

\begin{align}
\label{fullq}
&\frac{\p \delta_Q}{\p \tau} -3 (\omega-\hat{c}_s^2) \mathcal{H} \delta_Q + (1+\omega) \p_i v_Q^i  -   \frac{9\mathcal{H}^2}{\nabla^2}(1+\omega) (\hat{c}_s^2-\omega)\p_i v_Q^i\simeq \Theta(\delta^2),\nonumber\\
\nonumber\\
&v^{i\,\prime}_Q+ \mathcal{H}(1-3 \hat{c}_s^2)v^i_Q +\nabla_i \Phi +\frac{\hat{c}_s^2\,\partial_i  \delta_Q}{1+\omega} \simeq \Theta(\delta^2)\; ,
\end{align}
and the Poisson equation reads: 
\begin{align}
\label{fullP}
&\nabla^2 \Phi\simeq\frac{3}{2}\mathcal{H}^2\Omega_m \left(\delta_m + \frac{\Omega_q}{\Omega_m}\delta_Q \right)\equiv \frac{3}{2}\mathcal{H}^2\Omega_m \delta_T\; .\nonumber \\
\end{align}
A few comments are in order. We have restricted our analysis to linear order because, for the sake of consistency relations, we are mainly concerned with the behaviour of one long mode; for such mode the linear approximation is valid by definition. We have neglected a contribution proportional to $9\mathcal{H}^2/\nabla^2$ in the Poisson equation as this term is much smaller than unity for the scales of interest, well inside the horizon\footnote{Note that the same conclusions cannot \cite{Anselmi:2011ef} be drawn, at least at early times, for the similar term in the continuity Eq.~(\ref{fullq}).}. We have introduced the gauge-invariant quantity $\hat{c}_s^2$, the sound speed in the quintessence rest frame \cite{Bean:2003fb}. The non-linear corrections we are indicating as $\Theta(\delta^2)$ in Eqs.(\ref{fullm}-\ref{fullP}) would necessarily have to account also for corrections to $\hat{c}_s^2$. Finally, in the $\hat{c}_s^2\rightarrow 0$ limit one obtains the starting point for Eq.(\ref{packaged}), which is derived for $\delta_{T}$ as defined in the right hand side of Eq.~(\ref{fullP}) and where it has been assumed $v^i_Q=v^i_m\equiv v^i$. Linearly, both the reduced system in Eq.(\ref{packaged}) closed by the corresponding Poisson's equation and the full system in Eqs.(\ref{fullm}-\ref{fullP}) in the limit $\hat{c}_s\rightarrow 0$,  
are invariant the under time-dependent translations
\bea
\tau\rightarrow \tilde{\tau}=\tau\,;\, x^i \rightarrow \tilde{x}^i=x^i + n^i(\tau)\,;\, v_{m,Q}^i \rightarrow v_{m,Q}^i + n^{i '}\nonumber \\
\delta_{m,Q} \rightarrow \tilde{\delta}_{m,Q}=\de_{m,Q}\,; \,\Phi\rightarrow \tilde{\Phi}= \Phi-x^i(\H n^{i\prime}+n^{i\prime\prime})\,, \;
\label{fulltransf}\;
\eea

\noi  and this is precisely the (Galilean) symmetry that generates CRs in large scale structure. The effect of a non-zero $\hat{c}_s$ in the full system is intuitively clear: it renders the Euler equation (and solution) for the two species different already at the linear level. A different large scale limit for two velocities is a direct violation of the equivalence principle and leads to a modification of standard CRs. A measurement of a non-zero $\hat{c}_s$ is then a direct signature of CRs breaking. Let us describe more in detail how consistency relations emerge in a general set-up.

\subsection{Brief General Treatment}

One key ingredient for consistency relations to be in place is the existence of a field that transforms (also) non linearly under a residual diff we shall call $``s"$. In the following, up to and including Eq.~(\ref{less}), we will provide a qualitative presentation of how consistency conditions work in our and similar setups.
In coordinate space, the generic action of a gauge transformation ``$s$" on a field ``$\varphi$" is:
\bea
\delta_s \varphi \equiv i [Q_{s}, \varphi]=  \underbrace{(...)\varphi}_{\rm linear} + \underbrace{(...)}_{\rm non-linear}\; ,
\label{both}
\eea
where $Q_s$ is the charge associated with the symmetry $s$ and we have not specified the expression on the right hand side because it depends on the specific transformation ``s". The dots preceding $\varphi$ typically stand  for a differential operator and the only relevant information on the non-linear piece is that it is \textit{not} proportional to powers of $\varphi$\footnote{We refer the reader to \cite{Rizzo:2016akm} for a recent interesting application of CRs with the velocity field as soft mode. The work in \cite{Rizzo:2016akm} relies on the fact that, in our language, the velocity field has both linear and non-linear transformations.}. To give an example other than the transformation in Eq.~(\ref{fulltransf}), the action of a dilaton symmetry on a field whose transformation has also a non-linear component is \cite{Assassi:2012zq}:
\bea
\delta_d \zeta= i [Q_d, \zeta]=-1 - \vec{x}\cdot \partial_{\vec{x}}\zeta\; ,
\label{dilatation}
\eea

\noi where in this case the non-linearly transforming field is a gravitational degree of freedom, specifically the metric scalar perturbation $\zeta$. Let us now apply this reasoning on the transformation law for the 2-point function made up by two hard modes. From Eq.~(\ref{both}) it follows that:
\bea
\la [Q_{s}, \varphi\varphi] \ra \propto \frac{\partial}{\partial} \la\varphi\varphi \ra + \underbrace{...}_{\rm non-linear}, 
\label{4}
\eea
where we use the symbol $\frac{\partial}{\partial}$ to denote the generic differential operator we inherit from the linear transformation component of the field, such as the second term on the r.h.s. of Eq.~(\ref{dilatation}). We use the proportionality sign between left and right hand side in Eq.(\ref{4}) to keep the discussion general and stress again that, for the purposes of this brief discussion on the properties of consistency relations, it will not be necessary to specify the details of a particular ``s" transformation. The conclusions we will draw from Eq.~(\ref{4})-(\ref{less}) will be fairly general and only later adapted to the system under study. The unspecified non-linear terms on the r.h.s. of Eq.(\ref{4}) stand for pieces corresponding to diagrams which are {not \textit{connected}} in the standard field theory sense and therefore we omit them in what follows. The proof of this last statement for a system endowed with the symmetries in Eq.~(\ref{fulltransf}) can be found in \cite{Hinterbichler:2012nm}. There is another, equivalent, way to express the action of the symmetry via the charge $Q_s$, it relies on introducing a complete set of mutually orthogonal states \cite{Assassi:2012zq} (see also \cite{Dimastrogiovanni:2015pla}):
\bea
\la [Q_{s}, \varphi\varphi] \ra = 2i\, {\rm Im} \big[\la Q_{s} \sum_{n}( | n \ra \la n |)  \varphi\varphi \ra \big]\,.
\label{5}
\eea 
For the sake of simplicity and without loss of generality, let us limit here the number of such states to two: $| \varphi\ra$ and $| \sigma \ra $, which we assume to have been already \textit{orthogonalized}. Whenever there is an expectation value (such as the one that defines our chief observable, the total power spectrum) that satisfies conjugate symmetry, linearity and positive-definiteness, one may define a corresponding scalar product. It follows that orthogonalization procedures, such as the Gram-Schmidt method, can be applied.

\noi In practice, in this setup  $|\varphi\rangle$ stands as a placeholder for any of the fields in Eq.~(\ref{fulltransf}), that is $\delta_{m,Q}$, as well as  $\Phi$ and $v_i^{m,Q}$. As will be clear from Eq.(\ref{averaging}) however, only non-linearly transforming fields will give rise to non-trivial consistency relations.
Indeed, combining Eq.~(\ref{4}) and (\ref{5}), it follows that
\bea
\frac{\p}{\p} \la \varphi \varphi \ra \propto 2i\, {\rm Im} \big[\la Q_{s} | \varphi \ra \la \varphi |  \varphi\varphi \ra + \la Q_{s} | \sigma \ra \la \sigma |  \sigma\varphi \ra  \big].
\eea
If $\varphi$ and $\sigma$ transform only linearly under the symmetry, by virtue of the averaging process one has 
\bea
\la Q_{s} | \varphi \ra \propto \frac{\p}{\p} \la \varphi \ra = 0 =  \frac{\p}{\p}  \la \varphi,\varphi \ra \; ,
\label{averaging}
\eea
and no non-trivial consistency relation is active. If instead only e.g. $\varphi$ has a non-linear transformation component, then it follows that:

\bea
\la Q_{s} | \varphi \ra = c_{\rm number}(q\rightarrow 0) \Rightarrow  \frac{\p}{\p} \la \varphi \varphi \ra \propto  \la \varphi_{q\rightarrow 0} \,\varphi\varphi \ra \, ,
\eea
indicating that the CR holds in the simplest form. Finally, if also $\sigma$ has a non-linear component the result is a less standard: \bea
\frac{\p}{\p} \la \varphi \varphi \ra \propto  \la \varphi_{q\rightarrow 0} \,\varphi\varphi \ra + c_{\rm r}  \la \sigma_{q\rightarrow 0} \,\varphi\varphi \ra\, ,
\label{less}
\eea
where $c_{\rm r}$ stands for a relative coefficient, unimportant for the present discussion. If both terms in Eq.(\ref{less}) are non-zero one concludes that  
the squeezed contribution of quantities 
 such as  $\la \varphi_{q\rightarrow 0}\,  \varphi_{k} \varphi_{q-k} \ra$ may not be gauged away. This carries important observational consequences which are well known e.g. in the context of multi-field inflationary models where CRs breaking contains information on the mass, the spin, and the coupling of the extra\footnote{In the inflationary context, ``extra" is to be understood with respect to the inflaton field, whilst in our LSS set-up it refers to any species other than cold dark matter.} particles \cite{Kehagias:2015jha,Arkani-Hamed:2015bza}.

\subsection{Relevance for dark energy and modified gravity}

\noi Let us now apply this line of reasoning to the system in Eqs.(\ref{fullm}-\ref{fullP}).  From Eq.~(\ref{fulltransf}) it is clear that there are several modes that transform non-linearly under the symmetry: $\Phi, \pi_m,\pi_Q$, the last two being the velocity potentials (defined as $\partial_i\,\pi^{\,m,Q}\equiv v^{m,Q}_i $) associated to, respectively, $v_m$ and $v_Q$. Furthermore, the two velocities (potentials) have generically independent solutions. In the most general setup the action of a soft mode is then of the form in Eq.~(\ref{less}) and does therefore \textit{break} standard CRs. We stress that such dynamics is not at all limited to a quintessence component but applies to e.g. general dark energy and modified gravity dynamics.\\ 
\noi One may ask what happens in a reduced system such as that of Eq.~(\ref{packaged}). The latter obtained, as usual, by imposing on the full system that $\hat{c}_s\rightarrow 0$ and $v_m=v_Q$. The quintessence component precisely tracks dark matter in this limit. The assumption of a common velocity for all species is akin to having, intuitively speaking, one and a half degrees of freedom (this ``naive" counting gives two degrees of freedom only if both density and velocity fields are independent for dark matter and quintessence).  The consistency of this assumption has been demonstrated in earlier literature \cite{Creminelli:2009mu} and further confirmed in \cite{Sefusatti:2011cm,Anselmi:2011ef}. The reduced system enjoys invariance under Eq.(\ref{fulltransf}). There is more: only one velocity potential is now in play and its action as a soft mode is not independent from that of the Newtonian potential as one may readily verify from the linear eom. The two act in fact as \textit{alternative} soft ``pions" \cite{Horn:2014rta} and therefore deliver a standard consistency relation.\\ \noi It is instructive at this stage to point out yet another route to non-standard consistency relations in the case of modified gravity. Modification Einstein's equation in low density environments are allowed so long as general relativity (GR) is recovered in high density regions (e.g. the solar system) where observations are compatible only with very small deviations from $\Lambda$CDM. The transition from order one modification of the laws of gravity to GR  can occur by means of several screening mechanism (see \cite{Babichev:2013usa,Joyce:2014kja} for a review). Screening dynamics puts to the test one of the requirements in our CRs checklist: adiabaticity. For the standard relation to be in place, the equation of motion of the (one and only) gauge parameter describing the residual diff must be the same as that of a long physical mode.\\
Crucially, screening dynamics entails a solution e.g. for the density contrast field that changes according to its being in the (un)screened region. It is then impossible to identify a single time-dependent and spatially independent gauge mode $n(\tau)$ that always satisfies the adiabaticity condition. This immediately leads to modified CRs \cite{Kehagias:2013rpa}.

\section{Clustering of biased tracers}
\label{cbt}
\noi One may ask how the effects of additional dofs propagate all the way to biased tracers observables. Just as in $\Lambda$CDM, here too the effect of short distance physics on long wavelength dynamics is encoded in an effective stress-energy tensor. There is however also an effective force: it accounts for the momentum exchange (at short distances) between dark matter and additional species, mediated by gravity. One can {employ} the bias models developed in {\cite{McDonald:2009dh,Senatore:2014eva,Assassi:2014fva,Angulo:2015eqa}} to derive the results for two or more species. Due to the fact that the formation time of a collapsed object is approximately of order Hubble, one should account for the density of a given collapsed object to depend on the underlying history, in other words on long-wavelength fields evaluated over a length of time going back at least one Hubble time. This explains the integral over time in the following expression for the halo field:
\begin{align}
&\delta_h(\vec x,t)\simeq \int^t H(t')\; \left[  c_{\delta_T}(t')\; \frac{\delta_T(\vec x_{\rm fl},t')}{H(t')^2}+ c_{\delta_{\rm{d.e.}}}(t')\, \delta_{\rm{d.e.}}(\vec x_{\rm fl}) \right. \nonumber \\ 
&\quad+ c_{\partial v_{c}}(t')\, \frac{\partial_i v^i_c(\vec x_{\rm fl},t')}{H(t')}+ c_{\partial v_{\rm{d.e.}}}(t') \frac{\partial_i v^i_{\rm{d.e.}}(\vec x_{\rm fl},t')}{H(t')} \nonumber \\
&+ c_{\epsilon_c}(t')\,\epsilon_c(\vec x_{\rm fl},t')+ c_{\epsilon_{\rm{d.e.}}}(t')\, \epsilon_{\rm{d.e.}}(\vec x_{\rm fl},t') 
\left.+ c_{\partial^2\delta_T}(t')   \;\frac{\partial^2_{x_{\rm fl}}}{k_{\rm M}^2}\frac{\delta_T(\vec x_{\rm fl},t')}{H(t')^2}+
\ldots \right] \, ,
\label{bias}
\end{align}
where the dependence on the density fields and corresponding bias coefficients $c_{\delta_T},c_{\delta_{\rm{d.e.}}}$ as well as the dependence on the velocity fields and corresponding parameters $c_{\partial v_{c}}, c_{\partial v_{\rm{d.e.}}}$  are outlined. The stochastic component regulated by $ c_{\epsilon_c}, c_{\epsilon_{\rm{d.e.}}}$ is introduced to account for the difference between the average dependence of, for example, the galactic field on a given realization of the long wavelength dark matter fields, and its response in a specific realization.
One ought to also include in this framework the presence of a non-trivial length scale enclosing the radius of influence in the halo formation. This scale will be approximately of the same order as the range covered by the matter that ended up in a given collapsed object. The wavenumber corresponding to this scale is conventionally called $k_M$ and it accounts for the last term in the above formula. For a detailed derivation of Eq.~(\ref{bias}) we refer the reader to the work in \cite{Senatore:2014eva} for the single species case and  \cite{Angulo:2015eqa} for the two species (dark matter + baryons) generalization.
The definition of the flow variable is:
\bea
\vec{x}_{\rm fl}(\vec{x},\tau,\tau^{\prime})=\vec{x}- \int_{\tau^{\prime}}^{\tau}d \tau^{\prime\prime} \vec{v}(\tau^{\prime\prime},\vec{x}_{\rm fl}(\vec{x},\tau,\tau^{\prime}))\; .
\eea

\noi Note also that, in principle, the definition of the ``flow" variable, $\vec x_{\rm fl}$, accounting for the halo formation, can be different for different species \cite{Angulo:2015eqa}. \noi The above equation resembles the DM+baryons result, although there exists one important difference. The deviation from the EdS-like approximation {(i.e. assuming that kernels in Eq.~(\ref{eq:kernels}) are time-independent)} introduces new operators in the bias expansion above. Indeed, upon using Eq.~(\ref{bias}), one can show that the time dependence of the kernels in Eq.~(\ref{eq:kernels}), encoded in terms such as $\epsilon^{(1)},~\epsilon^{(2)},\nu_3$,  introduces (after formally performing the time integration) an independent bias coefficient (see \cite{Angulo:2015eqa}). At second order in the field $\delta_h^{(2)}$, this is not expected to yield any new independent bias operators due to degeneracies in the operators momentum dependence.
For $\delta_h^{(3)}$, on the other hand, we expect one new {independent} operator to arise due to the effects of time evolution. The same kind of reasoning holds already in $\Lambda$CDM cosmology (if one does \textit{not} assume the EdS-like approximation) but there the effects are known to be small. Crucially, the presence of additional dofs, such as dark energy,  will magnify them.

\section{Conclusions}

{\noi Given the vast amount of data that is being currently gathered by several astronomical surveys of the galaxy distribution, it is both important and  timely to study how the additional dynamics typical of beyond-$\Lambda$CDM models may affect large scale structure observables. We have provided one such study by focussing on the dynamics of a dark matter+clustering quintessence system.
In particular, we detailed on the properties of the power spectrum  and suggested a parallel with the behaviour of the non-equal time pure dark matter correlators. The non-trivial time dependence carried by the parameter $C$ in Eq.~(\ref{packaged}) is what enables such comparisons.\\
\noi  We then studied so called consistency relations  for a more general setup than the reduced dark matter+quintessence system by restoring a non zero sound speed for the latter and clarified under what conditions consistency relations are modified.
Dark energy and modified gravity models exhibit the characteristic features that prevent one from being able to ``gauge away" the squeezed contribution of observables such as the bispectrum. We show that, as one probes higher orders in perturbation theory, this conclusion holds true also at the level of biased tracers.\\ 
\noi Whilst the dynamics of the reduced system has been solved analytically to all orders\footnote{The caveat here being the addition of counterterms from the effective field theory treatment.} in perturbation theory, a detailed analysis of the observables from the full system requires considerably more numerical work, something that we plan to address in a forthcoming study \cite{toappear}. Another necessary next step is to investigate possible degeneracies between the effects on observables due to  extra dynamical degrees of freedom and those sourced by primordial non-Gaussian initial conditions, going beyond local type non-Gaussianity.\\

\noi \textit{Acknowledgements}\\
{\noi We are delighted to thank Lam Hui, Patrick McDonald, and Massimo Pietroni for illuminating discussions. MF is grateful to E. Dimastrogiovanni for insightful comments.
MF is supported in part by NSF PHY-1068380; ZV is supported in part by DOE contract  DEAC02-76SF00515.}

\section{Appendix}
\label{appe}
For a thorough derivation of the results reported here we refer the reader to \cite{Fasiello:2016qpn}. We start by reporting  the typical ansatz for the total density contrast solution:

\begin{align}
\delta^{}_{\bf k}(\eta)
&=\sum_{n=1}^{\infty} 
F^{}_n({\bf q}_1..{\bf q}_n,\eta) D_+^{n}(\eta)\delta_{\bf q_1}^{\rm  in}..\delta_{\bf q_n}^{\rm in}\; ,
\label{ansatz}
\end{align}
where $\delta_{{\bf q}}^{\rm in}$ represents the initial value of the total density contrast. One should stress that an explicit time dependence is directly present in the kernels $F_{n}$. The definition for the basic kernels $\alpha, \beta$ is as follows: 
\bea
&&\alpha(\vec{q}_1,\vec{q}_2) =1+({\bf q}_1\cdot {\bf q}_2)/{\bf q}_1^2\; , \qquad \beta(\vec{q}_1,\vec{q}_2) =({\bf q}_1+{\bf q}_2)^2 ({\bf q}_1\cdot{\bf q}_2) /2q_1^2 q_2^2\; ,
\eea
and the corresponding symmetrized and shortened expression for $\alpha$ kernels is: 
\begin{align}
\alpha^s_{12,3}(\vec{q}_1,\vec{q}_2,\vec{q}_3)=\frac{1}{2}\Big[\alpha(\vec{q}_1+\vec{q}_2,\vec{q}_3)+\alpha(\vec{q}_3,\vec{q}_1+\vec{q}_2) \Big]\, .
\end{align}

\noi The time-dependent kernels $\mu_n, \nu_v$ are angle-averaged quantities whose dynamics is regulated by the equations \cite{Bernardeau:2001qr}:
\begin{align}
\label{eq:average}
&\dot{\nu}_n + n\,\nu_n-\mu_n = \frac{1}{C} \sum_{m=1}^{n-1} {n \choose m} \mu_m \, \nu_{n-m}\, , \\
&\dot{\mu}_n + (n-1)\mu_n -\tfrac{f_-}{f_+^2}(\mu_n-\nu_n) = \frac{1}{3C} \sum_{m=1}^{n-1} {n \choose m} \mu_m \, \mu_{n-m}\,,\nonumber
\end{align}
with initial conditions $\nu_1=\mu_1=1$ and where $C$ is the time-dependent quantity defined below Eq.~(\ref{packaged}). Employing and generalizing the same procedure as in \cite{Bernardeau:2001qr}, in \cite{Fasiello:2016qpn} the following relations were derived for $\lambda_i$ and $\kappa_i$
\begin{align}
\label{eq:lambda_kappa}
&\dot{\lambda_i}+3 \lambda_i-\kappa_i=\frac{1}{C}\left(\nu_2\, c_{\lambda_i}^{\nu2}+\mu_2\, c_{\lambda_i}^{\mu2}\right)\; , \nonumber\\
&\dot{\kappa_i}+2 \kappa_i-\frac{f_-}{f^2_+}(\kappa_i-\lambda_i)= \frac{1}{C}\,\mu_2\, c_{\kappa_i}^{\mu2}\; ,
\end{align}
where the index $i$ can take values $\left\{1,2 \right\}$.
These equations, similarly to those for $\mu_n, \nu_n$, can be efficiently integrated numerically. 
Continuity with well-known results in Einstein-de Sitter space can be obtained by demanding in Eq.~\eqref{eq:lambda_kappa} above:
\bea
\label{eq:c_choice}
c_{\lambda_1}^{\nu2}=c_{\lambda_1}^{\mu2}=c_{\lambda_2}^{\nu2}=2 c_{\lambda_2}^{\mu2}=1\; , \qquad \; 
c_{\kappa_1}^{\mu2}=c_{\kappa_2}^{\mu2} = 0 \; .
\eea
The momentum dependence for the third order kernels is given by:
\begin{align}
\mathcal{F}_3^{\epsilon}&=-\frac{1}{12}\Bigg[(\alpha^{s}_{12,3}-3\beta_{12,3})(3\beta_{12}-\alpha^{s}_{12}) +  {\rm 2\, perm.}_{\rm cross}\Bigg]\; ,\nonumber\\
\mathcal{F}_3^{\nu_3}&=\frac{1}{8}\Bigg[\left(\, \alpha^s_{1,23}(\alpha^s_{23}-3\beta_{23})+\beta_{1,23}(\alpha^s_{23}+\beta_{23})\,\right)+   {\rm 2\, perm.}_{\rm cross}\Bigg]\; , \nonumber\\ 
\mathcal{F}_3^{\nu_2}&=\frac{1}{4}\Bigg[(\alpha^{s}_{12,3}-\beta_{12,3})(3\beta_{12}-\alpha^{s}_{12}) +  {\rm 2\, perm.}_{\rm cross}\Bigg]\; , \nonumber\\
\mathcal{F}_3^{\lambda_1}&=\frac{1}{16}\Bigg[\left( \,\alpha_{12,3}(3\alpha^s_{12}+7\beta_{12}) + \alpha_{1,23}(-9\alpha^s_{23}+19\beta_{23})-2  \beta_{1,23}(\alpha^s_{23}+9\beta_{23})\, \right) +   {\rm 2\, perm.}_{\rm cross}\Bigg]\; , \nonumber\\
\mathcal{F}_3^{\lambda_2}&=\frac{1}{4}\Bigg[ \left(\, \alpha_{1,23}(3\alpha^s_{23}-5\beta_{23}) - \alpha_{12,3}(\alpha^s_{12}+\beta_{12})-2 \beta_{1,23}(\alpha^s_{23}-3\beta_{23}) \,\right) +   {\rm 2\, perm.}_{\rm cross}\Bigg]\; ,
\end{align}
\noi where by \textit{cross} permutations in, for example, the quantity $\alpha_{1...m,m+1..n}$ it is meant those permutations that exchange momenta in the $(1....m )$ set with those in the $(m+1...n)$ set.

\end{document}